\documentclass[twocolumn,letterpaper,nofootinbib,showpacs,floatfix]{revtex4}
\usepackage{graphicx}
\usepackage{amsmath}
\usepackage{amsfonts}
\usepackage{bm}
\usepackage{color}
\usepackage{dcolumn}

\newcommand{\Deqn}[1]{{Eq.\ (\ref{#1})}}
\newcommand{\Deqns}[1]{{Eqs.\ (\ref{#1})}}
\newcommand{\lm}{{l m}}
\newcommand{\calE}{{\cal E}}
\newcommand{\calT}{{\cal T}}
\newcommand{\Lie}{{\cal L}}
\newcommand{\tot}{{\text{tot}}}
\newcommand{\ret}{{\text{ret}}}
\newcommand{\R}{{\text{R}}}
\renewcommand{\SS}{{\text{S}}}
\newcommand{\ubar}{{\bar u}}
\newcommand{\rdot}{{\dot R}}
\newcommand{\rddot}{{\ddot R}}

\newcommand{\rwh}{{{}_\text{rw}}h}
\newcommand{\lzh}{{{}_\text{lz}}h}

\begin{document}
\title{A consequence of the gravitational self-force for\\
           circular orbits of the Schwarzschild geometry}
\author{Steven Detweiler }
  \affiliation{Institute for Fundamental Theory, Department of Physics,
  University of Florida, Gainesville, FL 32611-8440}
  \email{det@ufl.edu}
\date{April 22, 2008} 

\begin{abstract}
A small mass $\mu$ in orbit about a much more massive black hole $m$ moves
along a world line that deviates from a geodesic of the black hole geometry
by $O(\mu/m)$. This deviation is said to be caused by the gravitational
self-force of the metric perturbation $h_{ab}$ from $\mu$. For circular
orbits about a non-rotating black hole we numerically calculate the
$O(\mu/m)$ effects upon the orbital frequency and upon the rate of passage
of proper time on the worldline. These two effects are independent of the
choice of gauge for $h_{ab}$ and are observable in principle.  For distant
orbits, our numerical results agree with a post-Newtonian analysis including
terms of order $(v/c)^6$.
\end{abstract}
 \maketitle

\section{Introduction and summary}
\label{introductionsummary}

\subsection{Gravitational waveforms, perturbation analysis and the self-force}
\label{motivation}

The push towards the detection of gravitational waves depends upon accurate
theoretical models of the sources. Numerical relativity has made great
progress in the past few years and now tracks black hole binary systems for
many orbits and down to the final coalescence and ring down. These recent
computational successes are joining the post-Newtonian analyses in providing
building blocks for the construction of gravitational wave templates.

A gap in the theoretical progress remains for binary systems with an extreme
mass ratio. A small black hole in a close orbit about a much larger one
might be too relativistic for post-Newtonian analyses. And the great
difference in length scales for the two black holes causes difficulty for a
full numerical relativity solution. This gap is the natural domain of
perturbation theory.

The state of the art in perturbative gravitational waveforms \cite{Drasco06}
has a small mass $\mu$ moving along a geodesic of the metric $g_{ab}$ of the
much larger black hole. The perturbed Einstein equations then determine the
metric perturbation $h_{ab}$ which contains the waveform. But, these
waveforms would be much improved if the influence of $h_{ab}$ back on the
motion of $\mu$ were part of the analysis. Then the dissipative effects of
radiation reaction, joining with conservative effects, would be reflected in
the waveform and would allow tracking $\mu$ for more orbits without losing
phase information as the inspiral proceeds.

The metric perturbation $h_{ab}$ causes the worldline of $\mu$ to be a
geodesic of $g_{ab}+h_{ab}$.
 Sometimes the same motion is described as non-geodesic in
$g_{ab}$ with the acceleration $a^a_{\text{SF}}$ being caused by the
\textit{self-force} from $h_{ab}$ acting on $\mu$.
 In this paper, we prefer the ``geodesic in $g_{ab}+h_{ab}$'' description
of the motion, and use the phrase ``self-force'' only in a generic way to
describe any and all of the $O(\mu)$ effects of $h_{ab}$ on the world-line
of $\mu$, even those which are unrelated to the acceleration
$a^a_{\text{SF}}$.

When the mass $\mu$ is modeled as a point particle, an attempt to calculate
any consequence of the self-force immediately bumps against a technical
difficulty. The metric perturbation $h_{ab}$ is singular at the location of
$\mu$, precisely where it must be evaluated. Some method of regularization
must be invoked to remove the singular feature. The formal aspects of this
regularization are now very well understood \cite{DeWittBrehme60, Mino97,
quinn-wald:97, DetWhiting03, poisson:03, BarackOri00, Barack00, BarackOri02,
BarackOri03, Mino02, BMNOS02, Det05}, and we give some details for our
particular application in Sec.\ \ref{numericaldetails}.

\subsection{Our limitations}
\label{ourlimitations}
 In this paper we limit our analysis to the self-force effects upon the
circular orbits of the Schwarzschild geometry. The numerical analysis is
straightforward, and the theoretical interpretation of the results is clear
and unambiguous. These simple orbits provide a convenient test bed for ideas
regarding the self-force.

Despite the emphasis of our opening paragraphs, this paper contains no new
information regarding the dissipative effects of radiation reaction on the
gravitational waveform for slow, circular inspiral into a non-rotating black
hole. In Appendix \ref{fluxintegral} we sketch a proof, based upon the form
of the perturbed Einstein equation, that the self-force removes energy
directly from the orbit of $\mu$, via \Deqn{dEdt}, at a rate that matches
the energy being radiated away in the wave zone and down through the event
horizon. This result is satisfying but not unexpected. At this order of
approximation, the dissipative effects of the self-force imply that the
orbital frequency $\Omega$ should change at a rate
\begin{equation}
  \frac{d\Omega}{dt} = \mu\frac{dE}{dt}
           \times\frac{d\Omega/dr}{\mu\,dE/dr} + O(\mu^2)
\label{dOmegadt}
\end{equation}
where $dE/dt$ is evaluated from \Deqn{dEdt}, and $\mu E$ is the $O(\mu)$
contribution to the mass-energy of the system as measured at a great
distance, which is equivalent to $\mu T^au_a +O(\mu^2)$ \footnote{This
equivalency follows from Zerilli's \cite{Zerilli70,Zerilli} analysis of
monopole perturbations of the Schwarzschild metric.}, with $T^a$ being the
timelike Killing vector and $u^a$ being the four velocity of $\mu$.
 Thus, for circular orbits of the Schwarzschild geometry the self-force
formalism based upon first-order perturbation theory adds formality but
yields nothing new about the evolution of the gravitational waveforms from
the dissipative part of the self-force. In fact, conservative self-force
effects on the values of $E$ and $\Omega$ for circular orbits are described
in Sec.\
\ref{quasicircular} but these are not sufficient for improving the
calculation of $d\Omega/dt$ from dissipation without at least also
determining the $O(\mu^2)$ part of $dE/dt$.  This result is common knowledge
within the self-force community \cite{Poisson04c} but appears not to have
spread much beyond that.

Non-circular orbits in the Schwarzschild metric and all orbits in the Kerr
metric contain substantial complications that we also do not address.

\subsection{Our results}
\label{ourresults}

In perturbative matters, we prefer to focus upon those observable quantities
which can be defined in a manner independent of the gauge choice for
$h_{ab}$. The angular velocity $\Omega$ and the time component of the four
velocity $u^t$ are two such quantities for $\mu$ in a circular orbit. Our
main computational result is determining the change in the relationship
between $\Omega$ and $u^t$ caused by the self-force. The coordinate radius
of the orbit is not included in this analysis because it depends upon the
gauge choice and has no inherent physical meaning.

First, we define  $R_\Omega$ via $\Omega^2 = m/R_\Omega^3$ as a natural
radial measure of the orbit which inherits the property of gauge independence
from $\Omega$.
  In Sec.\ \ref{gaugeindependence} we show that $u^t$ naturally separates into two
parts $u^t = {}_0u^t + {}_1u^t$, where each part is individually gauge
independent.
 Further the functional relationships between $\Omega$, ${}_0u^t$ and
$R_\Omega$ are identical to their relationships in the geodesic limit, $\mu
\rightarrow 0$, and show no effect from the self-force.

\begin{figure}
  \includegraphics[clip,scale=0.35]{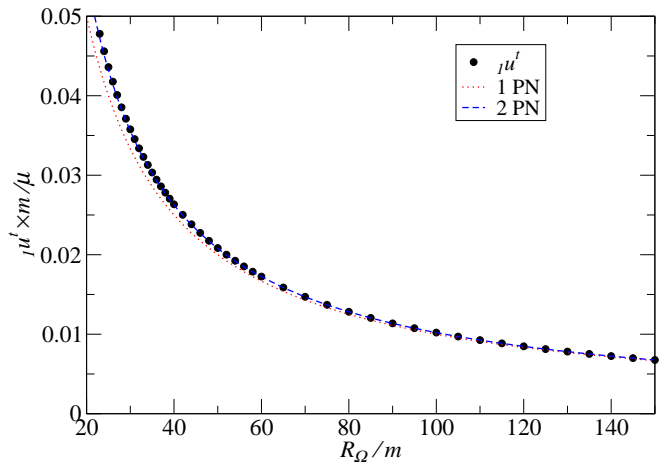} \caption{The quantity
  ${}_1u^t$, which is the gauge independent $O(\mu)$ part of $u^t$ effected
  by the self-force, is given as a function of $R_{\Omega}$ for circular
  orbits in the Schwarzschild geometry. Also shown are ${}_1u^t$ as
  calculated with 1PN and 2PN analyses of Appendix \ref{postnewtonian}
  based upon Refs.~\cite{BlanchetFayePonsot98} and \cite{BlanchetLR06}.
  }
\label{20-150uT}
\end{figure}

\begin{figure}
  \includegraphics[clip,scale=0.35]{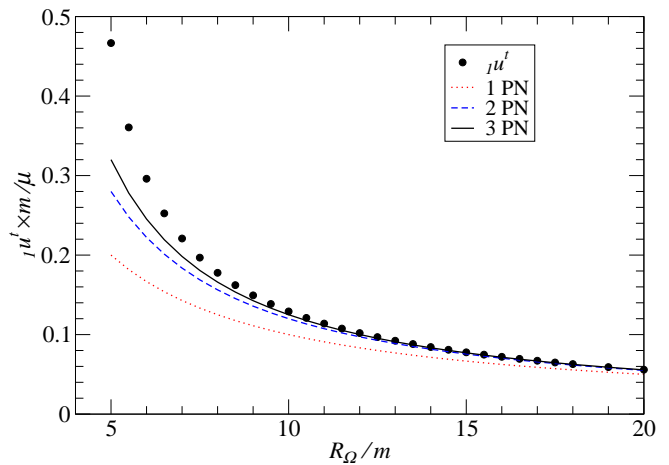}
  \caption{The same as Fig.\ \ref{20-150uT}, but including the 3PN
  analysis. }
\label{5-20uT}
\end{figure}

The remainder ${}_1u^t$ is, however, a true consequence of the self-force,
and we plot ${}_1u^t$ as a function of $R_\Omega$ in Figs.\ \ref{20-150uT}
and \ref{5-20uT}.
 These two figures along with Table \ref{data}
are the primary numerical results of this paper.


In Appendix \ref{postnewtonian} we derive a post-Newtonian (PN) expansion
for ${}_1u^t$ based upon the work of others \cite{BlanchetFayePonsot98,
BlanchetLR06}.  Our expansion is in powers of  $m/R_\Omega$, which is
$v^2/c^2$ in the Newtonian limit, and from \Deqn{PNexpanA}
\begin{equation}
  {}_1u^t = \frac{\mu}{m} \left[
    -  \left(\frac{m}{R_\Omega}\right)
    - 2 \left(\frac{m}{R_\Omega}\right)^2
    - 5 \left(\frac{m}{R_\Omega}\right)^3 + \cdots \right] ,
\label{PNexpan}
\end{equation}
which includes terms of order $v^6/c^6$.

Our Figs.~\ref{20-150uT} and \ref{5-20uT} display the different levels of
the post-Newtonian expansion. These agree with our numerical results as well
as they should.

With a good bit of hubris, we have gone one step further. After the three
known post-Newtonian terms were removed from our determined values of
${}_1u^t$, we numerically fit the residual to an expansion of higher order
terms, and found the coefficient of $({m}/{R_\Omega})^4$ to be $-27.61\pm.03$
\footnote{We had already used a similar technique while only knowing the
analytic expansion up to the $({m}/{R_\Omega})^2$ term. Finding the
coefficient of $({m}/{R_\Omega})^3$ to be $-5.001$, with the last digit
uncertain, provided the impetus to push the analytical work to one more
order.}.

We now describe an astronomical thought-observation. Imagine that an
observed system has a small mass in a circular orbit, with orbital frequency
$\Omega$, about a large non-rotating black hole. Assume that the mass $m$ of
the black hole could be determined by some independent means, and that a
redshift observation, perhaps as described in Sec.\ \ref{redshift}, could
determine $u^t$. Then
\begin{equation}
  {}_0u^t = [1-3(\Omega m)^{2/3}]^{-1/2}
\end{equation}
from \Deqn{GIuT},
\begin{equation}
  R_\Omega = \left(\frac{m}{\Omega^2}\right)^{1/3}
\end{equation}
from the definition of $R_\Omega$, and
\begin{equation}
  {}_1u^t = u^t - {}_0u^t .
\end{equation}
With values of $m$, $R_\Omega$ and ${}_1u^t$ in hand, an astronomer could
then determine the mass of the small object from the graphical data.

Our self-force result may only be of academic interest. But it is a result.
Our numerical calculations are consistent with a post-Newtonian expansion.
We estimate the unknown coefficient of $v^8/c^8$ in the expansion.  And we
have proposed one unlikely observational method to infer the mass of a small
object which is orbiting a much larger black hole.

\subsection{Outline}

In Sec.\ \ref{geodesicsofpertS} we introduce our notation, and we provide
the components of the geodesic equation for the perturbed Schwarzschild
metric $g_{ab}+h_{ab}$ in Sec.\ \ref{geodesicequation}.

In Sec.\ \ref{quasicircular} we focus on quasi-circular orbits---those that
would be circular except for the dissipative effects of the self-force.
 And we find that two components of the four-velocity, $u^t$ and $u^\phi$, are
effected by the self-force in a manner that is $O(\mu)$ and independent of
the particular gauge that is chosen for $h_{ab}$. Their ratio is the orbital
frequency $\Omega = u^\phi/u^t$, which is also gauge independent.

The component $u^t$ is the ratio of the rate of passage of Schwarzschild
time to proper time at the particle. Two different physical interpretations
of $u^t$ for quasi-circular orbits are given in Sec.\ \ref{interpretuT}; one
of these is in terms of a redshift measurement, the other is related to the
helical Killing vector field of the perturbed geometry.

The formal effect of the self-force on the relationship between $\Omega$ and
$u^t$ is described in Sec.\ \ref{gaugeindependence}. Particular emphasis is
placed upon how the gauge independence of this effect reveals itself.

Section \ref{numericaldetails} contains details of our numerical
determination of the relationship between $u^t$ and $\Omega$. Table
\ref{data} contains an extensive summary of our numerical results which
might be used by others for a numerical comparison. This section also has a
description of the regularization required by our application which points
out some difficulties inherent in regularizing any quantity which is not
gauge independent, such as the acceleration of $\mu$ as measured in the
background metric.

Our conclusion in Sec.\ \ref{conclusion} describes both the current state of
self-force calculations as well as its possible future.

We have relegated a number of side discussions to Appendices.

Some convenient identities are given in Appendix
\ref{elementary-identities}.

Gauge transformations that lead to the understanding that $u^t$, $u^\phi$
and $\Omega$ are gauge independent are given in
Appendix~\ref{gaugetransformations}.

An elementary, completely Newtonian example of the confusion that gauge
freedom can cause is in Appendix~\ref{newtoniangaugeproblem}.

In Appendix \ref{fluxintegral} we show that the rate at which $\mu$ loses
energy via the self-force precisely matches the rate that gravitational
waves carry energy out at infinity and down the black hole.

The post-Newtonian expansion for ${}_1u^t$ in \Deqn{PNexpan} is derived in
Appendix \ref{postnewtonian}.

\section{Geodesics of the perturbed Schwarzschild geometry}
  \label{geodesicsofpertS}
Self-force analysis implies that a point mass $\mu$ moves along a geodesic
of the perturbed Schwarzschild metric $g_{ab} + h_{ab}$ where, in this
section, $h_{ab}$ is the suitably regularized metric perturbation and is at
least ${\cal C}^1$.
 We assume that the geodesic is in the equatorial plane.
 We assume that $h_{ab}=O(\mu)$, and we work only through first order in $\mu$
or $h_{ab}$. Indices are lowered and raised with $g_{ab}+h_{ab}$ and
$g^{ab}-h^{ab}$.
 We use Schwarzschild coordinates, and $h_{ab}$ is a function of $t$, $r$,
$\theta$ and $\phi$.
  To be clear: we
specifically choose to normalize the four velocity $u^a$ of the particle
with respect to the actual regularized physical metric $g_{ab}+h_{ab}$, and
not with respect to the abstract background Schwarzschild metric $g_{ab}$.

The quantities $E$, $\dot R$, and $J$ are functions of the proper time $s$
and are defined in terms of the particle's four-velocity by
\begin{equation}
   E \equiv -u_t , \qquad \rdot \equiv u^r
                   \qquad\text{and}\qquad  J \equiv u_\phi,
\label{utpR}
\end{equation}
which leads to
\begin{equation}
 u_a = \left( -E, \frac{\dot R+ u^b h_b{}^r}{1-2m/r} , 0, J \right)
\label{ua}
\end{equation}
and
\begin{equation}
 u^a = \left( \frac{E+ u^b h_{bt}}{1-2{m} /{r}}, \dot R, 0,
                  \frac{J-u^b h_{b\phi}}{{r}^2}   \right),
\label{uA}
\end{equation}
where the overdot represents $d/ds$. Thus, $u^a$ depends upon $s$ implicitly
through its dependence upon $E$, $\dot R$, $J$ and also upon $r$, evaluated
at the particle.
 These expressions for $u_a$ and $u^a$ are recursive, but the recursion is
intended to be carried only through first order in $h_{ab}$.

It is convenient to define the non-radial part of the four-velocity of the
particle, with no $h_{ab}$ terms included, as
\begin{equation}
 \ubar^a \equiv \left( \frac{E}{1-2{m} /{r}},0, 0,
                  \frac{J}{{r}^2}   \right).
\label{ubar}
\end{equation}
Also, the instantaneous angular velocity of the particle is
\begin{equation}
  \Omega \equiv u^\phi/u^t .
\label{OmegaDef}
\end{equation}

\subsection{Geodesic equation}
\label{geodesicequation} One convenient exact form of the geodesic equation
for the metric $g_{ab}+h_{ab}$ is
\begin{equation}
  \frac{du_a}{ds} = \frac{1}{2} u^b u^c \frac{\partial}{\partial x^a}
       (g_{bc} + h_{bc}) .
\label{geodesic}
\end{equation}

While analyzing this geodesic equation we drop all terms of $O(\mu^2)$. The
components of \Deqn{geodesic} provide
\begin{eqnarray}
  \frac{dE}{ds} &=& -\frac{1}{2} u^a u^b \frac{\partial h_{ab}}{\partial t},
\label{dEds} \\
  \frac{dJ}{ds} &=& \frac{1}{2} u^a u^b \frac{\partial
  h_{ab}}{\partial\phi},
\label{dJds} \\
  \frac{d}{ds}\Big(\frac{{r}\rdot}{{r}-2{m} } + u^a h_{ar}\Big) &=&
     \frac{1}{2} u^a u^b \frac{\partial}{\partial r} ( g_{ab} + h_{ab} ) .
\label{Rddot}
\end{eqnarray}
These and other equations describing the particle's worldline are to be
evaluated at the location of the particle.

The normalization of $u^a$ is a first integral of the geodesic equation,
\begin{eqnarray}
 1 &=& -u^a u^b(g_{ab}+h_{ab})
\nonumber\\
   &=&  \frac{E^2 - \dot R^2}{1-2{m} /{r}} - \frac{J^2}{{r}^2}
             + {\ubar}^a {\ubar}^b h_{ab} -\dot R^2 h_{rr}
\label{norm}
\end{eqnarray}
which is derived by beginning with \Deqn{uA}.

\subsubsection{The radial component}

Further analysis of the radial component of the geodesic equation is lengthy
and requires care.
We begin with \Deqn{Rddot}, and use \Deqns{ua} and (\ref{uA}) extensively
along with the elementary identities in Appendix \ref{elementary-identities}
which allow for the easy commutation of $\ubar^a$ with $\partial_r$.
Simplification proceeds  by identifying all terms which are $O(\mu^2)$ if the
motion is quasi-circular as is described in Sec.\ \ref{quasicircular}; these
terms are systematically moved to the left hand side. Toward the end of the
analysis the normalization condition (\ref{norm}) is used to replace $E^2$ in
favor of $J^2$ on the right hand side. Final simplification arises from a
convenient placement of the indices of $h_{ab}$ on the left hand side. The
result is
\begin{eqnarray}
 \rddot (1+h_{r}{}^r) 
  \!\!&+&\!\!  \frac12 \rdot^2  \partial_r h_{r}{}^r
  + \ubar^a u^b \partial_a h_b{}^r 
  + \dot E h_{tr}
  + \dot J h^{\phi r} 
 \nonumber\\
  &=&  -\frac{{m}}{r^2}(1-{\ubar}^a {\ubar}^b h_{ab})
   + \frac{J^2}{r^4}(r-3{m})
\nonumber\\
  && {} + (1-2{m}/r)\frac12 \partial_r (\ubar^a \ubar^b h_{ab}) .
\label{RddotA}
\end{eqnarray}

\subsection{Quasi-circular orbits} \label{quasicircular}

For any orbit \Deqns{dEds} and (\ref{dJds}) imply that $\dot E \sim \dot J
\sim O(\mu)$.
 In this section our interest focuses on the orbits which would be circular
but for these effects of radiation reaction.
  For these \textit{quasi-circular} orbits we require that
$\dot R = O(\mu)$ and also that $\ddot R = O(\mu^2)$. The second of these
conditions is necessary to insure that the orbit is not a precessing
elliptical orbit with a very small eccentricity $\epsilon=O(\mu)$.
 To summarize, for a quasi-circular orbit  $E$, $R$ and $J$ change slowly,
$\dot E  \sim  \rdot \sim \dot J \sim O(\mu)$, and their rates of change are
slower still, $\ddot E \sim \rddot \sim \ddot J \sim O(\mu^2)$.

In this section we continue to drop $O(\mu^2)$ terms including those which
are consequences of the quasi-circular condition from the geodesic
equations. Thus the normalization of $u^a$ becomes
\begin{equation}
 1 = \frac{E^2}{1-2{m} /{r}} - \frac{J^2}{{r}^2}
      + {\ubar}^a {\ubar}^b h_{ab} .
\label{qcnorm}
\end{equation}

Neither $T^a \partial/\partial x^a = \partial /\partial t$ nor $\Phi^a
\partial/\partial x^a = \partial /\partial \phi$ is a Killing vector of
$g_{ab}+h_{ab}$, but a helical Killing vector $k^a$ exists for a
quasi-circular orbit in that $\Lie_k (g_{ab}+h_{ab}) =O(\mu^2)$. The Killing
vector exists for any gauge choice for $h_{ab}$. But if the gauge choice is
reasonable and $h_{ab}$ respects the helical symmetry, then $k^a$ is
particularly simple \cite{Det05} with
\begin{eqnarray}
  k^a \partial/\partial x^a &=& \partial /\partial t
         + \Omega\partial/\partial \phi,
\nonumber\\
 \Lie_k g_{ab} &=& 0, \text{ and } \Lie_k h_{ab} = O(\mu^2).
\label{Liekh}
\end{eqnarray}
In what follows, the only assumption which we make about the gauge is that
it respects both the helical symmetry over a dynamical timescale and also
the natural reflection symmetry through the equatorial plane. Our results
are independent of the assumption, but the analysis is much simplified
because of it. For example, ${\ubar}^a$ is tangent to a trajectory of $k^a$
so that ${\ubar}^a
\partial_a h_{bc} = O(\mu^2)$ at the particle for a quasi-circular orbit.

For quasi-circular orbits \Deqns{dEds} and (\ref{dJds}) simply imply that
\begin{eqnarray}
  \frac{dE}{dt} &=& -\frac{1}{2\ubar^t} \ubar^a \ubar^b\partial_t h_{ab}
  \quad\text{and}
\label{dEdt} \\
  \frac{dJ}{dt} &=& \frac{1}{2\ubar^t}\ubar^a \ubar^b \partial_\phi h_{ab}
\label{dJdt} ,
\end{eqnarray}
and \Deqn{Liekh} then yields
\begin{eqnarray}
  \frac{dE}{dt} = \Omega \frac{dJ}{dt} .
\end{eqnarray}
We show in Appendix \ref{gaugetransformations} that the right hand sides of
\Deqns{dEdt} and (\ref{dJdt}) are both independent of the gauge choice for
$h_{ab}$. Thus, $dE/dt$ and $dJ/dt$ are gauge independent for quasi-circular
orbits.

For a quasi-circular orbit every term on the left hand side of the radial
equation (\ref{RddotA}) is $O(\mu^2)$. It easily follows that
\begin{equation}
  \frac{J^2}{r^2} = \frac{{m}}{{r}-3{m}}(1-\ubar^a\ubar^bh_{ab})
  -\Big(\frac{{r}-2{m}}{{r}-3{m}}\Big)
       \frac{r}{2} \frac{\partial }{{\partial r}}
    \big( \ubar^a\ubar^b h_{ab} \big) .
\label{J2eq}
\end{equation}
\Deqns{qcnorm} and (\ref{J2eq}) together imply that
\begin{equation}
  \frac{E^2 r(r-3{m})}{(r-2{m})^2} =
      1 - \ubar^a\ubar^b h_{ab}
            - \frac{r}{2} \frac{\partial }{\partial r}\big(\ubar^a\ubar^b
            h_{ab}\big) .
 \label{E2eq}
\end{equation}
It is important to note that the $O(\mu)$ parts of $E$ and $J$ depend upon
the choice of gauge and have no precise physical meaning in terms of the
energy or angular momentum of the particle or the system. There is some
irony that $dE/ds$ and $dJ/ds$ are gauge independent, while $E$ and $J$ are
not.

Two interesting quantities for quasi-circular orbits are components of the
four velocity $u^a$,
\begin{eqnarray}
  (u^t)^2  &=& \Big(\frac{dT}{ds} \Big)^2  =
  \frac{(E + \ubar^b h_{tb})^2}{(1-2{m} /{r})^2}
\nonumber\\  &=&
  \frac{{r}}{{r}-3{m} }  \Big(1+\ubar^a \ubar^b h_{ab}
     - \frac{{r}}{2}\ubar^a \ubar^b\partial_r  h_{ab}\Big)
\label{uTeqn}
\end{eqnarray}
and
\begin{eqnarray}
  (u^\phi)^2 &=& \Big(\frac{d\Phi}{ds} \Big)^2  =
  \frac{1}{r^4} (J- \ubar^b h_{\phi b})^2
\nonumber\\  &=&
      \frac{({r}-2{m}) }{{r}({r}-3{m} )}
         \left[ \frac{{m} (1+\ubar^a \ubar^b h_{ab})}{{r}({r}-2{m})}
     - \frac{1}{2} \ubar^a \ubar^b\partial_r h_{ab} \right]. \qquad
\label{uPeqn}
\end{eqnarray}
These follow from \Deqns{E2eq} and (\ref{J2eq}) along with the equations in
Appendix \ref{elementary-identities}.

A gauge transformation is an infinitesimal coordinate transformation
$x^a_\text{new} = x^a_\text{old} + \xi^a$, which transforms
 $h_{ab}\rightarrow h_{ab}+\Delta h_{ab}$ where $\Delta h_{ab} = -2\nabla_{(a}\xi_{b)}$.
 For $h_{ab}$ respecting the symmetry of the helical Killing vector, we show in Appendix
\ref{gaugetransformations}
 that $\Delta (\ubar^a\ubar^b h_{ab}) = 0$ at $\mu$. Thus the right hand side
of \Deqn{uTeqn} changes only in two places. The change in $r/(r-3m)$
resulting from $r\rightarrow r+\xi^r$ is
\begin{equation}
 \Delta \left[\frac{r}{r-3m}\right] = -\frac{3m}{(r-3m)^2} \xi^r
\end{equation}
 and the change
\begin{equation}
  \Delta(\ubar^a\ubar^b\partial_r h_{ab}) = - \frac{6m}{r^2(r-3m)}\xi^r
\end{equation}
is from \Deqn{uudhgauge}. These two changes cancel each other at $O(\mu)$ on
the right hand side of \Deqn{uTeqn}, which is therefore gauge independent. A
similar argument shows that $u^\phi$ is also gauge independent.

Further, the ratio of $u^\phi$ and $u^t$ gives the orbital frequency
\begin{equation}
  \Omega^2 = \Big(\frac{u^\phi}{u^t}\Big)^2
     = \frac{{m} }{{r}^3}
        - \frac{{r}-3{m} }{2 {r}^2} \ubar^a \ubar^b \partial_r h_{ab}
\label{Omega2}
\end{equation}
for a quasi-circular orbit. Its gauge independence is inherited from $u^t$
and $u^\phi$, and may also be demonstrated directly following the same steps
as for $u^t$.

While $\Omega$ is observable from infinity and is gauge independent, the
Schwarzschild radial coordinate of the orbit is neither. A gauge
transformation simply changes the the radius of the orbit by $\Delta r =
\xi^r$.

\subsection{Physical interpretations of $u^t$} \label{interpretuT}

The quantity $u^t$ is the ratio of the rates of change of Schwarzschild
coordinate time and of proper time along the particle's geodesic in the
perturbed and regularized Schwarzschild metric.  We have two different
physical interpretations which give further meaning to $u^t$.

\subsubsection{$u^t$ as a constant of motion}

The perturbed metric $g_{ab}+h_{ab}$ has a helical Killing vector $k^a$.
With the reasonable gauge choice that $h_{ab}$ respects the helical symmetry
as in \Deqn{Liekh}, it follows that
\begin{equation}
  - k^a u_a = E - \Omega J
\end{equation}
is a constant of motion over a dynamical timescale. For a general orbit
\begin{eqnarray}
  (E - \Omega J) u^t &=& E u^t  - J u^\phi
\nonumber\\ &=&
  \frac{E (E + u^b h_{tb})}{1-2{m} /{r}}
  - \frac{J}{{r}^2} (J - u^b h_{\phi b})
\nonumber\\ &=&
  1 +  \dot R\ubar^a h_{ar} + \dot R^2 \Big(\frac{r}{r-2m}+h_{rr}\Big)
\end{eqnarray}
with \Deqns{uA} and (\ref{norm}) being used to derive the second and third
equalities, respectively. For a quasi-circular orbit, the right hand side of
the last line is $1 + O(\mu^2)$, and thus
\begin{equation}
    E - \Omega J = 1/u^t.
\end{equation}
The gauge independence of $u^t$ then implies that the constant of the motion
$E - \Omega J$ is also gauge independent even while $E$ and $J$ are not.

\subsubsection{ $u^t$ as a redshift measurement}
 \label{redshift}

In principle $u^t$ might be measured by a redshift observation, which we now
describe.

Let a light source be located at the particle where the perturbed metric
is suitably regularized. Let $\nu^a$ be the tangent vector to an affinely
parameterized null geodesic of a photon from this light source. The energy
$\calE_\text{em}$ of the photon, as emitted, is proportional to
$u^a_\text{em} \nu_a$ with $u^a_\text{em}$ the four velocity of the
emitter. Thus the ratio of the energies as measured when observed and when
emitted is
\begin{equation}
   \frac{\calE_{\text{ob}}}{\calE_{\text{em}}}
        = \frac{u^a_{\text{ob}} \nu_a}{u^a_{\text{em}} \nu_a}
\end{equation}
with $u^a_\text{ob}$ the four velocity of the observer. With $k^a$ being
the helical Killing vector field of the perturbed metric, $k^a \nu_a$ is a
constant of motion along the geodesic of the photon. The four velocity of
the emitter is the four velocity of the particle, and $u^a_{\text{em}}
\propto k^a$ so that $u^a_{\text{em}} = u^t k^a|_{\text{em}}$. Let the
photon be observed along the $z$-axis at a large distance away from the
black hole so that $u^a_\text{ob} = u^t_\infty\delta^a_t=\delta^a_t$. We
then have
\begin{eqnarray}
   \frac{\calE_{\text{ob}}}{\calE_{\text{em}}}
  &=& \frac{u^t_\infty\nu_t^\infty}{u^t (k^a \nu_a)_{\text{em}}}
  = \frac{\nu_t^\infty}{u^t (k^a \nu_a)^{\infty}} .
\end{eqnarray}
Finally, this may be written as
\begin{eqnarray}
   \frac{\calE_{\text{ob}}}{\calE_{\text{em}}}
 &=& \frac{\nu_t^\infty}{u^t (\nu_t^\infty + \Omega \nu_\phi^{\infty})}
  = \frac{1}{u^t} - \frac{\Omega \nu_\phi^\infty}
                         {u^t (\nu_t^\infty + \Omega \nu_\phi^{\infty})}
\nonumber\\
  &=& \frac{1}{u^t}
\end{eqnarray}
because $\nu_\phi^\infty = 0$ at a large distance along the $z$-axis. Thus,
the gauge independent $1/u^t$ determines the redshift of a photon, emitted
from the particle, when the photon is observed on the $z$-axis at a large
distance.

\section{Gauge independence of our results}
\label{gaugeindependence}

In our analyses, we have a gauge independent expression for the orbital
frequency in \Deqn{Omega2},
\begin{equation}
 \Omega = \sqrt{\frac{m}{r^3}}
    \left[ 1 - \frac{r(r-3m)}{4m}\ubar^a\ubar^b\partial_r h_{ab} \right]
\label{GIomdef}
\end{equation}
which allows us to define a related gauge independent measure of the
separation between $m$ and $\mu$ via
\begin{equation}
  R_\Omega \equiv (m/\Omega^2)^{1/3}.
\label{Romdef}
\end{equation}
In the geodesic limit, $\mu\rightarrow0$, the separation $R_\Omega$ is
precisely the Schwarzschild radial coordinate of the orbit.

Our other gauge independent quantity of interest is $u^t$, in \Deqn{uTeqn},
which we now choose to write as
\begin{equation}
  u^t = {}_0u^t + {}_1u^t
\end{equation}
where
\begin{equation}
  {}_0u^t \equiv (1-3m/r)^{-1/2}
   \left(1 - \frac{r}{4} \ubar^a\ubar^b\partial_r h_{ab}\right)
\label{0uTdef}
\end{equation}
and
\begin{equation}
  {}_1u^t \equiv (1-3m/r)^{-1/2}\frac{1}{2}\ubar^a\ubar^b h_{ab} .
\label{1uTdef}
\end{equation}
From the discussions above and in Appendix \ref{gaugetransformations}, it is
clear that each of these parts of $u^t$ are individually gauge independent.

For the moment, assume that the analysis is being performed in a special
gauge where $\ubar^a\ubar^b\partial_r h_{ab} = 0$; the gauge vector required
to put a general $h_{ab}$ into this special gauge is given in
\Deqn{uudhgauge}.  In this case
\begin{equation}
  \Omega = \sqrt{\frac{m}{r^3}}, \quad R_\Omega = r
\label{omegaAndR}
\end{equation}
and
\begin{equation}
  {}_0u^t = (1-3m/r)^{-1/2} .
\label{0uT}
\end{equation}
 \Deqns{omegaAndR} and (\ref{0uT}) are equivalent to their geodesic limits.
It follows that, for this particular choice of gauge,
\begin{eqnarray}
  {}_0u^t &=&(1-3m/R_\Omega)^{-1/2}
\nonumber\\ &=&
         [1-3(\Omega m)^{2/3}]^{-1/2}
\label{GIuT}
\end{eqnarray}
Every bit of this equation is gauge independent. Therefore \Deqn{GIuT} must
hold for any choice of gauge. An alternative, direct verification of this
equation results from the use of
\Deqns{GIomdef}--(\ref{0uTdef}) to write both sides of \Deqn{GIuT} in terms
of $r$, $m$, $\ubar^a$ and $h_{ab}$, followed by the expansion of each side
through linear order in $h_{ab}$.

From our perspective the relationships between $\Omega$, $R_\Omega$ and
${}_0u^t$ are as if the motion were geodesic in $g_{ab}$ and independent of
the self-force and should be treated as such.

To see the effect of the self-force on $u^t$, we numerically determine
$\ubar^a\ubar^b h_{ab}$ and, thence, ${}_1u^t$ as a function of $R_\Omega$
as described in the next section. We have no need to evaluate
$\ubar^a\ubar^b\partial_r h_{ab}$.

Note that if the motion of $\mu$ were being described as ``accelerated in
the Schwarzschild geometry'', then $u^a$ would be normalized by $g_{ab}$.
Also $u^t$ would \textit{not} be gauge independent when defined in that
manner and the gauge independence of $u^t$ with our normalization from
$g_{ab}+h_{ab}$ would likely be undiscovered.

\section{Determination and regularization of $h_{ab}$}
\label{numericaldetails}

The major results or our numerical analysis were displayed in Figs.\ 1 and 2
of Sec.\ \ref{ourresults}. In this section we fill in some details regarding
how these results were obtained.

 \begin{table*}[h]
 \centering
 \begin{tabular}{r  l  l  l  l  l }
 \hline\hline
  $R_\Omega/m$\!\!\! & \quad${}_0u^t$ & \quad${}_1u^tm/\mu$ & $\quad-$3PN\quad  & \quad$\ubar^a\ubar^bh^\R_{ab}/\mu$\quad & \quad$\mu^{-2}dE/dt$ \\
 \hline
 $  5$ \em& $1.581138$ & $   -0.4666524$ & $     -0.14665$ & $   -0.5902738$ & $  0.002792737$ \\ 
 $  6$ \em& $1.414213$ & $   -0.2960275$ & $    -0.050657$ & $   -0.4186461$ & $ 0.0009403393$ \\
 $  7$ \em& $1.322875$ & $   -0.2208475$ & $    -0.022596$ & $   -0.3338900$ & $ 0.0004001632$ \\
 $  8$ \em& $1.264911$ & $   -0.1777197$ & $    -0.011704$ & $   -0.2809995$ & $ 0.0001961045$ \\
 $  9$ \em& $1.224744$ & $   -0.1493606$ & $   -0.0066994$ & $   -0.2439048$ & $ 0.0001059332$ \\
 $  10$ \em& $1.195228$ & $   -0.1291222$ & $   -0.0041222$ & $   -0.2160628$ & $ 6.151631\times10^{-5}$ \\
 $  12$ \em& $1.154700$ & $   -0.1019355$ & $   -0.0018198$ & $   -0.1765575$ & $ 2.429170\times10^{-5}$ \\
 $  14$ \em& $1.128152$ & $  -0.08438195$ & $  -0.00092714$ & $   -0.1495932$ & $ 1.115762\times10^{-5}$ \\
 $  16$ \em& $1.109400$ & $  -0.07205505$ & $  -0.00052185$ & $   -0.1298991$ & $ 5.708047\times10^{-6}$ \\
 $  18$ \em& $1.095445$ & $  -0.06290189$ & $  -0.00031616$ & $   -0.1148426$ & $ 3.166609\times10^{-6}$ \\
 $  20$ \em& $1.084652$ & $  -0.05582771$ & $  -0.00020271$ & $   -0.1029412$ & $ 1.871470\times10^{-6}$ \\
 $  25$ \em& $1.066003$ & $  -0.04359984$ & $  -7.98\times10^{-5}$ & $  -0.08180055$ & $ 6.157919\times10^{-7}$ \\
 $  30$ \em& $1.054092$ & $  -0.03577831$ & $  -3.75\times10^{-5}$ & $  -0.06788457$ & $ 2.486475\times10^{-7}$ \\
 $  35$ \em& $1.045825$ & $  -0.03034064$ & $  -1.99\times10^{-5}$ & $  -0.05802241$ & $ 1.155513\times10^{-7}$ \\
 $  40$ \em& $1.039750$ & $  -0.02633967$ & $  -1.15\times10^{-5}$ & $  -0.05066538$ & $ 5.950154\times10^{-8}$ \\
 $  50$ \em& $1.031421$ & $  -0.02084465$ & $  -4.65\times10^{-6}$ & $  -0.04041928$ & $ 1.962457\times10^{-8}$ \\
 $  60$ \em& $1.025978$ & $  -0.01724759$ & $  -2.22\times10^{-6}$ & $  -0.03362174$ & $ 7.926444\times10^{-9}$ \\
 $  70$ \em& $1.022142$ & $  -0.01470964$ & $  -1.19\times10^{-6}$ & $  -0.02878197$ & $ 3.681881\times10^{-9}$ \\
 $  80$ \em& $1.019294$ & $  -0.01282296$ & $  -6.94\times10^{-7}$ & $  -0.02516046$ & $ 1.894535\times10^{-9}$ \\
 $  90$ \em& $1.017095$ & $  -0.01136531$ & $  -4.32\times10^{-7}$ & $  -0.02234857$ & $ 1.054112\times10^{-9}$ \\
 $  100$ \em& $1.015346$ & $  -0.01020528$ & $  -2.82\times10^{-7}$ & $  -0.02010207$ & $ 6.238203\times10^{-10}$ \\
 $  120$ \em& $1.012739$ & $ -0.008475251$ & $  -1.35\times10^{-7}$ & $  -0.01673728$ & $ 2.515767\times10^{-10}$ \\
 $  140$ \em& $1.010889$ & $ -0.007246793$ & $  -7.30\times10^{-8}$ & $  -0.01433745$ & $ 1.167029\times10^{-10}$ \\
 $  160$ \em& $1.009508$ & $ -0.006329388$ & $  -4.27\times10^{-8}$ & $  -0.01253953$ & $ 5.998161\times10^{-11}$ \\
 $  180$ \em& $1.008438$ & $ -0.005618167$ & $  -2.66\times10^{-8}$ & $  -0.01114230$ & $ 3.334080\times10^{-11}$ \\
 $  200$ \em& $1.007585$ & $ -0.005050642$ & $  -1.74\times10^{-8}$ & $  -0.01002523$ & $ 1.971428\times10^{-11}$ \\
 $  250$ \em& $1.006054$ & $ -0.004032327$ & $  -7.13\times10^{-9}$ & $ -0.008016120$ & $ 6.476438\times10^{-12}$ \\
 $  300$ \em& $1.005037$ & $ -0.003355744$ & $  -3.43\times10^{-9}$ & $ -0.006677846$ & $ 2.607338\times10^{-12}$ \\
 \hline\hline                           
 \end{tabular}
 \caption{Summary of the gravitational self-force effects ${}_1u^t/\mu$, 
 $\ubar^a\ubar^bh^\R_{ab}$ and $dE/dt$
 for a variety of radii, $R_\Omega$. The label $-$3PN denotes
 the remainder after the post-Newtonian expansion in \Deqn{PNexpan} has been
 removed from ${}_1u^t$.} 
 \label{data}
 \end{table*}

\subsection{The retarded field}
\label{retardedfield}

We numerically determine the retarded metric perturbation $h^\ret_{ab}$ in
the vicinity of a small mass $\mu$ in a circular orbit about a large black
hole of mass $m$.  We use a standard frequency domain formalism that has
been in continuous use for nearly forty years. This formalism is based upon
a decomposition of the components of $h^\ret_{ab}$ in terms of tensor
harmonics. And we rely heavily upon the results of Regge and Wheeler
\cite{ReggeWheeler} and of Zerilli \cite{Zerilli70,Zerilli}. We use the
Regge-Wheeler gauge for $h^\ret_{ab}$, although all of our results are
independent of the gauge choice.

\subsection{Regularization}
\label{regularizationdetails}

Two different approaches to regularization have been developed for
self-force problems involving point particles and gravitational fields and
both lead to the same conclusion. Mino, Sasaki and Tanaka \cite{Mino97}
follow the DeWitt-Brehme \cite{DeWittBrehme60} analysis for electromagnetic
fields in curved spacetime and show that the gravitational self-force may be
described completely as geodesic motion in the perturbed geometry
$g_{ab}+h^{\text{tail}}_{ab}$, where $h^{\text{tail}}_{ab}$ is the part of
the retarded metric perturbation which comes from the support of the Green
function within the past null cone of the particle. Quinn and Wald
\cite{quinn-wald:97} invoke a physically appealing ``comparison axiom'' and
arrive at essentially the same conclusion.

A third approach \cite{DetWhiting03} notes that the tail part of the metric
perturbation is only a portion of a solution of the perturbed Einstein
equations. However a Green function, different from the retarded Green
function, determines the singular part of the metric perturbation
$h^\SS_{ab}$ which exerts no force on the particle. The remainder of the
metric perturbation $h^\R_{ab} \equiv h^\ret_{ab} - h^\SS_{ab}$ is regular
at the particle and is a source-free solution of the perturbed Einstein
equations. The self-force is subsequently described as geodesic motion in
the combined metric $g_{ab}+h^\R_{ab}$. At first perturbative order
$g_{ab}+h^\R_{ab}$ is a vacuum solution of the Einstein equations, which has
the desirable implication that a local observer would see the particle move
along a geodesic of a vacuum solution of the Einstein equations and, in
fact, would only observe the combined field $g_{ab}+h^\R_{ab}$ and have no
local method for distinguishing $h^\R_{ab}$ from $g_{ab}$.

A practical method for applying the regularization formalism was described
first by Barack and Ori \cite{BarackOri00,Barack00,BarackOri02,%
BarackOri03}, later by Mino, Nakano and Sasaki \cite{Mino02,BMNOS02} and
subsequently extended by others \cite{Lousto00,burko:00c,BarackLousto02,%
DetMessWhiting03,HaasP2006,Haas2007}.
  In this procedure the multipole moments of $h^\SS_{ab}$ and its derivatives
are calculated analytically and referred to as \textit{regularization
parameters}. The sum of these moments diverges when evaluated at the
particle, but each individual moment is finite. By construction $h^\SS_{ab}$
precisely matches the singularity structure of the retarded field at the
particle. Thus the difference of the retarded moments and the singular
moments gives a multipole decomposition of $h^\R_{ab}$ which converges when
summed over $l$ and $m$. Schematically, this procedure gives
\begin{equation}
  h^\R_{ab} = \sum_{\lm} h^{\R(\lm)}_{ab}
   = \sum_{\lm} \left[ h^{\ret(\lm)}_{ab}
        - h^{\SS(\lm)}_{ab} \right]
\label{modesum}
\end{equation}
for the regular field.

We show in Sec.\ \ref{gaugeindependence} that our problem only requires the
regularization of $\ubar^a \ubar^b h_{ab}$. Following the original
prescription of Barack and Ori \cite{BarackOri00} and extending it as in
Ref.~\cite{DetMessWhiting03}, we first evaluate
\begin{equation}
 \ubar^a \ubar^b h^{\ret(l)}_{ab}
    \equiv \sum_{m=-l}^{l} \ubar^a \ubar^b h^{\ret(lm)}_{ab} .
\end{equation}
Then we use the ansatz that
\begin{eqnarray}
  \ubar^a \ubar^b h^{\SS (l)}_{ab} &=&
        B + \frac{C}{l+1/2}  -  \frac{D}{(2l-1)(2l+3)}
  \nonumber\\
    && {} +  \frac{E_1}{(2l-3)(2l-1)(2l+3)(2l+5)}
  \nonumber\\
    && {} + O(l^{-6})
\label{FABDE}
\end{eqnarray}
where $B$, $C$, $D$, $E_1, \ldots$ are the regularization parameters. The
particular $l$ dependence of the $D$, and $E_1, \ldots$ terms is related to
the expansion of $(1-\cos\theta)^{n+1/2}$ in terms of Legendre polynomials
$P_l(\cos\theta)$; details are derived and described in Appendix D of Ref.
\cite{DetMessWhiting03}.

For our problem it is known analytically that $C=0$ and
\begin{equation}
  B = 2\sqrt{\frac{r-3m}{r^2(r-2m)}}\;
  {}_2F_1\left(\frac{1}{2},\frac{1}{2},1,\frac{m}{r-2m}\right)
\end{equation}
where ${}_2F_1$ is a hypergeometric function, and $r$ is the Schwarzschild
radial coordinate of the circular orbit. Work in progress will describe the
derivation of a variety of regularization parameters, including these. This
knowledge of $B$ and $C$, but not $D$, implies that
\begin{equation}
  \ubar^a \ubar^b h^{\R}_{ab} = \sum_{l=0}^\infty
  ( \ubar^a \ubar^b h^{\ret (l)}_{ab} - \ubar^a \ubar^b h^{\SS (l)}_{ab} )
\end{equation}
converges as $1/l$. To increase the rate of convergence, we augment our
knowledge of $B$ and $C$ by numerically determining further regularization
parameters \cite{DetMessWhiting03}: We use the fact that the behavior of
$\ubar^a \ubar^b h^{\ret(l)}_{ab}$, evaluated at $\mu$, must match $\ubar^a
\ubar^b h^{\SS (l)}_{ab}$ as given in \Deqn{FABDE} for large $l$. This
allows us to fit the numerical data to determine the additional
regularization parameters $D$, and $E_1$ to $E_3$. This results in the sum
converging as $\sim l^{-9}$. We sum up to $l=40$, providing sufficient
accuracy for the results as presented.

A second gauge independent quantity is $dE/dt$ as given in \Deqn{dEdt}. The
regularization parameters for this quantity are all zero, no regularization
is required, and the sum over $l$ converges faster than any power of $l$.

Table \ref{data} has a complete set of the interesting, gauge-independent
data regarding a circular orbit of a Schwarzschild black hole. Any other
gauge independent quantity, known by us, may be derived from these data. The
results for ${}_1u^t$ with the post-Newtonian expansion are displayed in
Figs.\ \ref{20-150uT} and \ref{5-20uT} of Sec.\ \ref{introductionsummary}.

Barack and Sago \cite{BarackSago2007} have recently used a time-domain
formalism to calculate the actual gravitational self-force in the Lorenz
gauge. Their $F_r$ corresponds to our
$\frac{1}{2}\mu\partial_r\big(\ubar^a\ubar^b h^\R_{ab}\big)$. However, we
worked in the Regge-Wheeler gauge and are not able to compare results.

\subsection{Gauge difficulties with regularization parameters}

If the singular and retarded fields in \Deqn{modesum} are in different gauges
then the schematic description of regularization fails.

The singular field is commonly described in the Lorenz gauge
$\lzh^\SS_{ab}$, while the retarded field is easily found in the
Regge-Wheeler \cite{ReggeWheeler} gauge $\rwh^\ret_{ab}$, in the context of
the Schwarzschild geometry. By gauge choice $\rwh_{t\phi}=0$ for even-parity
perturbations, while $\lzh_{t\phi}$ is generally not zero and is singular at
$\mu$. In this case $\rwh^\ret_{t\phi} - \lzh^\SS_{t\phi}$ is necessarily
singular, and the ``regularization'' procedure fails.

We now show that this failure is circumvented when a gauge independent
quantity is calculated.

Assume that the gauge vector relating the two gauge choices is known for the
retarded field,
\begin{equation}
  2\nabla_{(a} \xi_{b)} = \rwh^\ret_{ab} - \lzh^\ret_{ab},
\end{equation}
and is used to change the retarded field into the Lorenz gauge mode by mode,
\begin{align}
 \lzh^{\R(\lm)}_{ab}
     &= \rwh^{\ret(\lm)}_{ab} - 2\nabla_{(a} \xi^{(\lm)}_{b)}  - \lzh^{\SS(\lm)}_{ab}
\nonumber\\
     &= \lzh^{\ret(\lm)}_{ab} - \lzh^{\SS(\lm)}_{ab} .
\end{align}
Each component of $\lzh^{\R}_{ab}$ is now regular and the sum over modes
converges.

In this paper, we are eager to calculate the gauge independent combination
$\ubar^a\ubar^bh^\R_{ab}$ for quasi-circular orbits which involves the sum
over multipole terms such as
\begin{align}
 \ubar^a\ubar^b\lzh^{\R(\lm)}_{ab}
    = \ubar^a\ubar^b & \rwh^{\ret(\lm)}_{ab}
         - 2\ubar^a\ubar^b\nabla_{(a} \xi^{(\lm)}_{b)}
\nonumber\\ &
         - \ubar^a\ubar^b \lzh^{\SS(\lm)}_{ab} .
\label{gaugesum}
\end{align}
The demonstration that $\ubar^a\ubar^bh_{ab}$ is gauge independent for
quasi-circular orbits does not rely on $h_{ab}$ being the retarded, the
singular or the regular field, and it clearly shows that
$\ubar^a\ubar^b\nabla_{(a} \xi^{(\lm)}_{b)} = 0$ at the particle, for any
gauge vector\footnote{The helical symmetry need not be respected by $\xi_a$,
but the entire discussion of the gauge independence is then substantially
more complicated.}. Thus the $\nabla_a\xi_b$ term in \Deqn{gaugesum} is
identically zero, and
\begin{equation}
  \ubar^a\ubar^bh^\R_{ab}
   = \sum_{\lm} \left[ \ubar^a\ubar^b \rwh^{\ret(\lm)}_{ab}
        - \ubar^a\ubar^b \lzh^{\SS(\lm)}_{ab} \right]
\label{GIsum}
\end{equation}
converges to a value that is independent of any gauge choice.

This technique can be used to regularize any gauge-independent linear
combination of components of $h_{ab}$ and its derivatives. In fact, Moncrief
\cite{Moncrief74a} noted that the non-zero components of $\rwh_{ab}$ may be
described in terms of gauge independent, linear combinations of components of
$h_{ab}$ and its derivatives in a generic gauge. Thus, if the regularization
parameters for $\rwh^\SS_{ab}$ are known and if $h^\ret_{ab}$ is known in an
arbitrary gauge, then what might be termed the regularized gravitational
self-force in the Regge-Wheeler gauge could be determined in just this
manner. And the result would, at least, be mathematically well defined.

\section{Conclusion}
\label{conclusion}

In Ref.~\cite{barack-ori:01} Barack and Ori state ``The meaningful
description of the gravitational self-force must include both
$F^\alpha_{\text{self}}$ and the metric perturbation $h_{\alpha \beta}$.''
We agree wholeheartedly with this statement. But we go a step further and
believe that the value in calculating the gravitational self-force, in any
particular gauge, is to apply it to a question whose answer is related to
some physical observable. And a physical observable ought to be independent
of the gauge choice. In Appendix \ref{newtoniangaugeproblem} we give an
example of how easily gauge confusion appears even in Newtonian physics
where the direction of the gravitational self-force contains no useful
physical information without additional knowledge about the coordinates and
how the self-force is being evaluated.

Similarly, we prefer to describe the effects of the self-force as geodesic
motion in the perturbed and regularized metric $g_{ab}+h_{ab}$.  The
alternative description ``acceleration $a^a_{\text{SF}}$ in $g_{ab}$''
depends upon the gauge choice and bears no relationship to any actual
acceleration which an observer moving with the particle would measure with a
collection of small masses and springs. At this level of approximation $\mu$
is in free-fall in the actual, physical spacetime metric $g_{ab}+h_{ab}$. To
give $a^a_{\text{SF}}$ in some gauge seems to imply that $a^a_{\text{SF}}$
contains a true physical consequence---if that is the case we would prefer a
description of that consequence.

The circular orbits of the Schwarzschild metric perhaps provide the simplest
framework for any self-force problem. The angular decomposition and
elementary discrete frequency spectrum imply that only ordinary differential
equations need to be solved in order to determine the metric perturbations.
Apparently, few problems can be formulated in a gauge independent way within
this simple framework. We would be surprised if another first-order
self-force effect in this particular context were found which did not have a
solution in terms of the data available in Table I.

A general orbit has none of the natural symmetry of the quasi-circular orbit.
Then none of our $\Omega$, $u^t$, $u^\phi$, $E$ or $J$ can be described in
terms of gauge independent quantities. In this situation, where understanding
the meaning of a question is likely to be as difficult as determining its
answer, perhaps the only significant questions concern the waveform.

However, our preference for Gauge invariance is a matter of taste. So it is
important to note that
the value of $a^a_{\text{SF}}$ in the Lorenz gauge for circular orbits is
precisely defined and can be determined in a precise way, as ably
demonstrated by Barack and Sago \cite{BarackSago2007}. Their time-domain
implementation is sharply focused on $a^a_{\text{SF}}$ and takes a
significant step in the direction of calculating the elusive self-force
effect upon waveforms.

There is reason for optimism regarding waveforms. 
The difficulty of two disparate length scales for the extreme-mass-ratio
problem can be avoided by using the analytically known singular field
$h^\SS_{ab}$ to construct a smoothed out source \cite{BarackGoldbourn2007,
BarackSago2007, BarackGS2007, VegaDet08} for the wave equation of an
approximation to the regular field $h^\R_{ab}$. The smoothed out source shows
no structure with a length scale of $\mu$. The numerically determined
$h^\R_{ab}$ can be guaranteed to be at least $C^2$ at $\mu$ \cite{Det05,
VegaDet08}. In this case $h^\R_{ab}$ directly gives the self-force effects on
the motion and simultaneously provides the waveform in the wave zone.

To take full advantage of the self-force formalism will require that
$h^\R_{ab}$ be evaluated at second-order in $\mu$. But, this impediment does
not appear to be fundamental.
 It is likely that techniques based upon the current successes of numerical
relativity will be able finally to reveal a better description of the
dissipative effects of the self-force on gravitational waveforms.

\acknowledgments
 The author is grateful for discussions with Leor Barack, Luc Blanchet,
Eric Poisson, Norichika Sago, Ian Vega and Bernard Whiting.

Development of the ideas in this paper began while the author was at the
Aspen Center for Physics during the 2005 summer workshop LISA Data: Analysis,
Sources, and Science; we gratefully acknowledge the Center and the workshop
organizers for their support and kind hospitality. These ideas were further
developed at the eighth, ninth and tenth Annual Capra meetings at the
Rutherford Appleton Laboratory Oxford (2005), University of Wisconsin in
Milwaukee (2006) and University of Alabama in Huntsville (2007),
respectively. And we are grateful to the organizers of these valuable
workshops. The author also acknowledges the University of Florida
High-Performance Computing Center (URL: http://hpc.ufl.edu) and the Institute
for Fundamental Theory (URL: http://www.phys.ufl.edu/ift) for providing
computational resources and support that have contributed to the research
results reported in this paper. This work was supported in part by the
National Science Foundation, grant No. PHY-0555484.

\appendix

\section{Convenient identities for geodesics of the perturbed
Schwarzschild geometry} \label{elementary-identities}

The following identities are useful for simplifying an equation by
transforming terms involving $E$ and $J$ back and forth into terms only
involving $\ubar^a$. These equations are all elementary consequences of
\Deqn{ubar}:
\begin{eqnarray}
  \ubar^a \ubar^b h_{ab} &=& \frac{rE}{r-2{m}} \ubar^b h_{tb}
                            + \frac{J}{r^2} \ubar^b h_{\phi b},
\end{eqnarray} \begin{eqnarray} 
  \frac{\partial\ubar^a}{\partial r} \ubar^b h_{ab} &=& -\frac{2mE}{(r-2{m})^2}
\ubar^b h_{tb}
                            - \frac{2J}{r^3} \ubar^b h_{\phi b},
\end{eqnarray}
and
\begin{eqnarray}
  \frac{2}{r}\ubar^a \ubar^b h_{ab}
    +  \frac{\partial \ubar^a}{\partial r} \ubar^b h_{ab}
       &=& \frac{2(r-3{m})}{(r-2{m})^2} E \ubar^b h_{tb},
\end{eqnarray} \begin{eqnarray} 
  \frac{2{m}}{r(r-2{m})}\ubar^a \ubar^b h_{ab}
    &+&  \frac{\partial \ubar^a}{\partial r} \ubar^b h_{ab}
\nonumber\\
      &=& -\frac{2(r-3{m})}{r^3(r-2{m})} J \ubar^b h_{\phi b} .
\end{eqnarray}

\section{Gauge transformations}
\label{gaugetransformations}

The change in the metric perturbation under a gauge transformation
$x^a_\text{new} = x^a_\text{old} + \xi^a$ where $\xi^a$ is considered to
be infinitesimal (also known as an infinitesimal coordinate
transformation) is
\begin{eqnarray}
  \Delta h_{ab} = - \Lie_\xi g_{ab}
    &=& - \xi^c \partial_c g_{ab} - g_{ac} \partial_b \xi^c
                                       - g_{cb} \partial_a \xi^c
\nonumber\\ &=& \nabla_a \xi_b - \nabla_b \xi_a.
\end{eqnarray}

To preserve reflection symmetry we assume that $\xi^\theta$ and its
derivatives are all zero on the equatorial plane. Then, for the Schwarzschild
geometry we obtain
\begin{equation}
  \Delta h_{rr} = \frac{2m}{(r-2m)^2} \xi^r - \frac{2}{(1-2m/r)} \partial_r
  \xi^r,
\end{equation}
\begin{equation}
  \Delta h_{tr} = - \partial_r \xi_t + \frac{2m}{r(r-2m)} \xi_t
                       - \frac{1}{1-2m/r} \partial_t \xi^r,
\end{equation}
\begin{equation}
  \Delta h_{tt} = \frac{2m}{r^2} \xi^r - 2 \partial_t \xi_t,
\end{equation}
\begin{equation}
  \Delta h_{\phi\phi} = - 2r \xi^r - 2 \partial_\phi
  \xi_\phi,
\end{equation}
\begin{equation}
  \Delta h_{t\phi} = - \partial_t\xi_\phi - \partial_\phi\xi_t,
\end{equation}
and
\begin{equation}
  \Delta h_{r\phi} =  - \partial_r\xi_\phi + \frac{2}{r} \xi_\phi
               - \frac{1}{1-2m/r} \partial_\phi\xi^r.
\end{equation}

For quasi-circular orbits of perturbed Schwarzschild we consider $h_{tt} +
\Omega h_{\phi t}$ and $h_{t \phi} + \Omega h_{\phi \phi}$; these are
proportional to $\ubar^ah_{at}$ and $\ubar^ah_{a\phi}$ respectively. Under
a gauge transformation
\begin{eqnarray}
 \Delta h_{tt} + \Omega \Delta h_{\phi t}
     &=&
          \frac{2m}{r^2} \xi^r
        - 2 \partial_t \xi_t
        - \Omega \partial_t\xi_\phi
        - \Omega\partial_\phi \xi_t ,\qquad
\label{htt+hpt}
 \\ \Delta h_{t\phi} + \Omega \Delta h_{\phi \phi}
   &=&
        - \partial_t  \xi_\phi
        - \partial_\phi \xi_t
        - 2\Omega r \xi^r
        - 2\Omega \partial_\phi \xi_\phi .\qquad
\label{htp+hpp}
\end{eqnarray}
Combining these first provides
\begin{eqnarray}
  \lefteqn{\Delta(h_{tt} + 2\Omega h_{t\phi} + \Omega^2 h_{\phi\phi})
    =  2r\Big(\frac{{m}}{r^3}-\Omega^2\Big) \xi^r }&&
\nonumber\\ &&
 \qquad {}  - 2(\partial_t+\Omega\partial_\phi) \xi_t
    - 2\Omega (\partial_t+\Omega\partial_\phi) \xi_\phi,
\end{eqnarray}
and subsequently
\begin{eqnarray}
  \ubar^a \ubar^b \Delta h_{ab} &=&
    \frac{E^2}{(1-2m/r)^2}\Delta(h_{tt} + 2\Omega h_{t\phi}
              + \Omega^2 h_{\phi\phi})\qquad\;
\nonumber\\
  &=& \frac{E^2}{(1-2m/r)^2} \bigg[2r\Big(\frac{{m}}{r^3}-\Omega^2\Big) \xi^r
\nonumber\\
  &&{}  - 2(\partial_t+\Omega\partial_\phi) \xi_t
    - 2\Omega (\partial_t+\Omega\partial_\phi) \xi_\phi \bigg] .
\label{uuDelh}
\end{eqnarray}
With the assumption that $\xi^a$ respects the helical symmetry $\Lie_k
\xi^a = (\partial_t+\Omega\partial_\phi) \xi^a = 0$, the above effect of a
general gauge transformation simplifies to
\begin{eqnarray}
  \ubar^a \ubar^b \Delta h_{ab} &=&
    \frac{E^2}{(1-2m/r)^2}
      \bigg[2r\Big(\frac{{m}}{r^3}-\Omega^2\Big) \xi^r \bigg]
\nonumber\\  & = & 0, \label{uuDelh_2}
\end{eqnarray}
when evaluated at the particle where $\Omega^2={m}/r^3 + O(h)$. Equation
(\ref{uuDelh_2}) and $\Delta\ubar^a=O(h)$ then imply that $\Delta(\ubar^a
\ubar^b
 h_{ab}) = 0$, at $O(h)$. The fact that
\begin{equation}
\Delta\left(\ubar^a \ubar^b
 \partial_t h_{ab}\right) = 0,
\end{equation}
at the particle, follows simply from \Deqn{uuDelh}. A similar argument
reveals the gauge independence of $\ubar^a \ubar^b \partial_\phi h_{ab}$ when
evaluated at the particle.

It also follows from \Deqn{uuDelh} that
\begin{eqnarray}
   \ubar^a \ubar^b \frac{\partial}{\partial r} \Delta h_{ab} &=&
  \frac{E^2}{(1-2m/r)^2}
           \Big(\frac{-6{m}}{r^3}\Big) \xi^r
\end{eqnarray}
evaluated at the particle. 
For a quasi-circular orbit $E^2/(1-2m/r)^2 =1/(1-3m/r)+O(\mu)$, so that a
general gauge transformation induces a change at the particle
\begin{equation}
  \Delta(\ubar^a \ubar^b \partial_r h_{ab})
      = -\frac{6{m} }{{r}^2({r}-3{m})} \xi^r .
\label{uudhgauge}
\end{equation}
This result shows that the expression for $\Omega^2$ in \Deqn{Omega2} is
invariant under a general gauge transformation, while the radius of the
orbit changes with the radial coordinate, $r_{\text{new}} = r_{\text{old}} +
\xi^r$, and is \textit{not} gauge independent.

\section{Gauge dependence of the Newtonian gravitational self-force}
\label{newtoniangaugeproblem}

In Newtonian gravity, when one mass $m_1$ of a circular binary is
infinitesimal, then the orbital frequency is $\Omega^2 = Gm_2/r^3$.
 If $m_1$ is small but finite, then $\Omega$ changes by $O(m_1/m_2)$.
This change in $\Omega$ is properly described as a consequence of a
Newtonian gravitational self-force \cite{DetPoisson04}.

The orbital acceleration of $m_1$ in a binary is given by
\begin{equation}
  r_1 \Omega^2 = \frac{G m_2}{(r_1+r_2)^2}
\label{Kepler}
\end{equation}
where $r_1$ and $r_2$ are the radii of the orbits and are also the distances
from the masses to the center of mass. Thus $m_1 r_1 = m_2 r_2$.

Note that
\begin{equation}
 \lim_{m_1\rightarrow0} r_1+r_2 = r_1 .
\end{equation}
And in this same limit,
\begin{equation}
 \lim_{m_1\rightarrow0} r_1\Omega^2 = G m_2/r_1^2.
\end{equation}
But how does the acceleration of $m_1$ change as the limit is taken? The
answer would determine the sign of the self-force.
 For an extreme mass ratio, the magnitude of the acceleration of $m_1$ is
precisely
\begin{equation}
  r_1 \Omega^2 = \frac{G m_2}{ (r_1+r_2)^2}
\label{ngsf+}
\end{equation}
on the one hand, but also
\begin{equation}
  r_1\Omega^2 = \frac{G m_2}{(r_1+r_2)^2} = \frac{G m_2}{r_1^2}
        \left(1- \frac{2 m_1}{m_2} + \ldots\right),
\label{ngsf-}
\end{equation}
on the other.

The acceleration and the self-force depend upon the detail of how $r_1+r_2$
approaches $r_1$.
 If the distance $r_1+r_2$ between the masses is held fixed during the limit
($r_1$ grows slightly while $r_2$ shrinks) then \Deqn{ngsf+} implies that
the acceleration of $m_1$ is constant and that there is no self-force.
 However, if $m_1$'s orbital radius $r_1$ is held fixed during the limit, then
\Deqn{ngsf-} implies that the limit of the acceleration is approached from below,
and the self-force points outward.

Even in Newtonian physics the gravitational self-force is ambiguously defined
and not particularly relevant to understanding \Deqn{Kepler}.

This Newtonian ambiguity is precisely equivalent to the gauge ambiguity of
the gravitational self-force in general relativity. For an
extreme-mass-ratio binary, if the origin of the Schwarzschild coordinates is
at the center of the black hole then the coordinate value of $r$ at the
orbit would represent the distance between the particle center and the black
hole center, i.e. $r_1+r_2$, and \Deqn{ngsf+} would imply that there is no
self-force. However, a dipole gauge transformation can move the origin of
the coordinates to the center of mass of the system and then the coordinate
value of $r$ at the orbit would represent the distance between the particle
and the center of mass, i.e. $r_1$ alone, and \Deqn{ngsf-} would imply that
the self-force points outward. Consequently, a simple dipole gauge
transformation substantially changes the appearance of the gravitational
self-force.

The resolution of this confusion in Newtonian physics is to refrain from
introducing the concept of the self-force. It seems reasonable that general
relativity would follow the Newtonian lead.

\section{Using the energy flux for radiation reaction}
  \label{fluxintegral}

For the special case of quasi-circular inspiral, it has long been common
knowledge that the dissipative effects of radiation reaction could be
calculated by determining (\textit{i}) the total energy of particle $\mu E$,
(\textit{ii}) the orbital frequency $\Omega$, and (\textit{iii}) the rate of
energy loss $dE_\tot/dt$ via gravitational waves. Each of these is to be
calculated as a function of the radius $r$ of the orbit of $\mu$. As energy
is lost $\mu$ slowly spirals inward, and the rate of change of $\Omega$ is
then expected to be
\begin{equation}
  \frac{d\Omega}{dt}
    = \frac{dE_\tot}{dt} \times\frac{d\Omega/dr}{\mu \,dE/dr}.
\label{dOmegadtAPP}
\end{equation}
In this Appendix we show that $d\Omega/dt$ in this equation is equivalent to
$d\Omega/dt$ as determined in \Deqn{dOmegadt}, which is based upon the
self-force formalism.

Equation (\ref{dOmegadtAPP}) and \Deqn{dOmegadt} appear similar but have an
important difference. In \Deqn{dOmegadtAPP}, ${dE_\tot}/{dt}$ is the rate
that energy is radiated out at a large distance and down into the black
hole, while in \Deqn{dOmegadt} ${\mu\,dE}/{dt}$ is determined locally from
$h_{ab}$ via \Deqn{dEdt} and is the rate at which the the self-force removes
energy from the particle. These two equations give the same $d\Omega/dt$
only if the rate that energy is lost through gravitational waves is equal to
the rate that the self-force removes energy from the orbit. We now show that
this is the case.


We assume that $g_{ab}$ is the Schwarzschild metric, and only in this
appendix an overdot represents a derivative with respect to the Schwarzschild
time coordinate $t$.

The perturbed Einstein equations, with a perturbing stress-energy tensor
$T_{ab}=O(\mu)$ being the source, may be written as (see Eq.~(35.58) of Ref.
\cite{MTW})
\begin{equation}
  16\pi T_{ab} = - E_{ab}(h) 
\label{EabTab}
\end{equation}
where
\begin{eqnarray}
  E_{ab}(h) &=& \nabla^2 h_{ab} + \nabla_a \nabla_b h
           - 2 \nabla_{(a}\nabla^c h_{b)c}
\nonumber\\ & &
      + 2{R_a}^c{}_b{}^d h_{cd}  
      + g_{ab} ( \nabla^c\nabla^d h_{cd} - \nabla^2 h ) ,
\label{Eab}
\end{eqnarray}
with $h \equiv h_{ab} g^{ab}$.

For arbitrary symmetric tensors $k^{ab}$ and $h_{ab}$, the operator
$E_{ab}(h)$ satisfies the identity \cite{Det05}
\begin{equation}
  k^{ab} E_{ab}(h) =
    \nabla_c F^c(k,h) - \big<k^{ab}, h_{ab}\big> 
\end{equation}
with
\begin{eqnarray}
  F^c(k,h) &\equiv&
  \bar{k}^{ab} \nabla^c \bar{h}_{ab}
    - \frac{1}{2} \bar{k} \nabla^c  \bar{h}
       - 2 \bar{k}^{cb} \nabla^a \bar{h}_{ab}
\label{Fdefn}
\end{eqnarray}
where $\bar{h}_{ab} \equiv h_{ab} - \frac12 g_{ab} h^c{}_c$, and similarly
for $\bar k_{ab}$. Also
\begin{eqnarray}
  \big<k^{ab}, h_{ab}\big> & \equiv &
        \nabla^c\bar k^{ab} \nabla_c \bar h_{ab}
      - \frac{1}{2}\nabla^c \bar k \nabla_c \bar h
 \nonumber\\ &&
  {} - 2 \nabla_a \bar k^{ac} \nabla^b\bar h_{bc}
  {} - 2 \bar k^{ab} {{{{R_a}^c}_b}^d} \bar h_{cd}
\end{eqnarray}
is symmetric under interchange of $h_{ab}$ and $k_{ab}$.

Let $k_{ab} = \dot h_{ab}$. Then a consequence of \Deqn{EabTab} is
\begin{equation}
    16\pi T^{ab} \dot h_{ab} =
     - \nabla_c F^c(\dot h,h) + \big<\dot h^{ab}, h_{ab}\big>
\label{A8}
\end{equation}
We now evaluate an integral of both sides of this equation over a constant
$t$ surface $\Sigma$.

The stress-energy tensor for a point mass moving along a path through space
$X^i(t)$ is
\begin{equation}
  T^{ab} = \mu\frac{u^au^b}{u^t\sqrt{-g}}\, \delta^3(x^i-X^i(t)) .
\end{equation}
So, an integral of the left hand side of \Deqn{A8} is
\begin{align}
   16\pi & \int_\Sigma T^{ab} \dot h_{ab} r^2\sin\theta \,dr\,d\theta\,d\phi
  = {}  \frac{16\pi\mu}{u^t} u^au^b\dot h_{ab},
\label{A10}
\end{align}
where the right hand side is evaluated at the particle.

Assume that $\Sigma$ is bounded by one two-sphere $\partial\Sigma_\infty$ at
a large radius and by a second two-sphere $\partial\Sigma_{2m}$ close to the
event horizon. And integrate \Deqn{A8} over $\Sigma$. Then substitute
\Deqn{A10} into the left hand side and the result is
\begin{align}
  16\pi\mu  \frac{u^au^b\dot h_{ab}}{u^t}
     =&
  - \int_\Sigma  \nabla_c F^c(\dot h,h) r^2 \sin\theta\,
     dr\,d\theta\,d\phi
 \nonumber\\ &
    + \frac12 \frac{d}{dt} \int_\Sigma \big< h^{ab}, h_{ab}\big>
                r^2\sin\theta \,dr\,d\theta\,d\phi
\end{align}
where the time derivative was moved outside the second integral by the
virtues of the time independence of the metric and of the symmetry of
$\big<,\big>$. That same term is then zero because $\mu$ is moving along a
circular geodesic and $h_{ab}$ respects the helical symmetry $\Lie_k
h_{ab}=0$. Thus, the integral is
\begin{align}
   16\pi\mu  &\frac{u^au^b\dot h_{ab}}{u^t}
   = - \int_\Sigma \partial_c \big[r^2 \sin\theta F^c(\dot h,h)\big] \,dr\,d\theta\,d\phi
\nonumber\\
   & = - \frac{d}{dt}\Big[\int_\Sigma  F^t(\dot h,h)\,r^2 \sin\theta
         \,dr\,d\theta\,d\phi \Big]
\nonumber \\ &
  \hskip.5in {} - \int_\Sigma \frac{\partial}{\partial x^i}
            \big[r^2 \sin\theta F^i(\dot h,h) \big] \,dr\,d\theta\,d\phi
\nonumber\\
    & {} = - \Big[\oint_{\partial\Sigma_\infty} - \oint_{\partial\Sigma_{2m}}\Big]
            \big[r^2 \sin\theta F^r(\dot h,h) \big] \,d\theta\,d\phi
\label{A12}
\end{align}
where the first term on the right hand side of the second equation is zero
because of the helical symmetry, and the third equality follows from Gauss'
law.

The definition of $F^a(k,h)$ in \Deqn{Fdefn} gives
\begin{eqnarray}
  F^r(\dot h,h) &=& \dot{\bar{h}}^{ab} \nabla^r \bar{h}_{ab}
     - \frac{1}{2} \dot{\bar h} \nabla^r \bar{h}
      {} - 2 \dot{\bar h}^{rb} \nabla^a \bar{h}_{ab}.\quad
\end{eqnarray}
Note that when the integrals in the last line of \Deqn{A12} are evaluated
close to the event horizon and far into the wave zone $F^r(\dot h,h)$ is
then equal to $32\pi {\calT_t}^r$, a component of the effective
stress-energy tensor of a gravitational wave in a background geometry as
given in Eq.~(35.70) of Ref.\ \cite{MTW}.

Thus, the right hand side of \Deqn{A12} is $32\pi$ times $dE_\tot/dt$, the
rate that the gravitational waves carry energy out through the boundaries of
$\Sigma$. Our conclusion is
\begin{equation}
  \frac{dE_\tot}{dt} = - \frac{\mu}{2 u^t} u^a u^b \partial_t h_{ab}
                     = \frac{dE}{dt} ,
\label{A17}
\end{equation}
so that \Deqn{dOmegadtAPP} and \Deqn{dOmegadt} have the same implications.

In a numerical study of quasi-circular orbits, a comparison of $dE_\tot/dt$
with $\mu dE/dt$ is a useful test of the numerical implementation and
accuracy, but is not a test of the self-force formalism.

For completeness, we note that the actual, slow inspiral of a quasi-circular
orbit is solely a result of the radiative energy and angular momentum losses
which lead to $\mu$ slowly falling inward to a tighter orbit. The radial
component of the self-force does no work on $\mu$ and is not responsible for
the inspiral.

 \section{The post-Newtonian expansions of $u^t$ and $\Omega$}
\label{postnewtonian}

Recently we have used the analysis by Blanchet, Faye and Ponsot (BFP)
\cite{BlanchetFayePonsot98, BlanchetLR06} to determine the post-Newtonian
relationship between the two gauge independent quantities $\Omega$ and $u^t$
in the extreme mass ratio limit.

We use many of the results and much of the notation of BFP without extensive
clarification, and we limit our interest to orbits which are circular up to
the effects of radiation reaction. In this appendix only, $c$ is \textit{not}
set to unity, and an expression of the form $O(n)$ refers to general terms of
order $1/c^{n}$ in the limit $c\rightarrow\infty$.

The two components of the binary system are ${m_1}$ and ${m_2}$, and
\begin{equation}
 {M} \equiv {m_1}+{m_2}
\end{equation}
is the total mass. The components are located at $\vec y_1$ and $\vec y_2$
with velocities $\vec v_1 = d\vec y_1/dt$ and $\vec v_2 = d\vec y_2/dt$,
whose magnitudes $v_1$ and $v_2$ are given in \Deqns{v1eq} and (\ref{v2eq})
below. For convenience the Cartesian dot product between two vectors is
denoted
\begin{equation}
 (v_1 v_2) \equiv \vec v_1 \cdot \vec v_2 = - v_1\, v_2
\end{equation}
for the circular orbits of interest.

Also the relative position of the masses is $\vec r_{12} \equiv \vec y_1 -
\vec y_2$, and the relative velocity is $\vec v_{12} \equiv \vec v_1 -
\vec v_2$. The magnitude of the relative velocity is
\begin{equation}
  v_{12} = \Omega r_{12}+ O(6)
\end{equation}
where $\Omega$ is the angular velocity. The $O(6)$ term is a consequence of
radiation reaction at 2.5PN.

We summarize some results from BFP and Blanchet's Living Review
\cite{BlanchetLR06}.
 One dimensionless measure of the strength of the gravitational field is
\begin{equation}
 \gamma \equiv \frac{G{M}}{r_{12}c^2} ,
 \label{gammadef}
\end{equation}
and a second is
\begin{equation}
   \frac{v_{12}^2}{c^2} = \frac{\Omega^2r_{12}^2}{c^2} + O(8).
\end{equation}
These two parameters are precisely equal only in the Newtonian limit,
\begin{equation}
  \gamma  = \frac{v_{12}^2}{c^2} +O(4).
\end{equation}
It is natural, then, to introduce a third dimensionless parameter
\begin{equation}
     x^3 \equiv \gamma^2 
        \frac{v_{12}^2}{c^2}
     = \gamma^2
       \frac{\Omega^2r_{12}^2}{c^2}
     =  \frac{G^2{M}^2\Omega^2}{c^6}
\label{xdef}
\end{equation}
which has the useful features that $x=v_{12}^2/c^2+O(4)$ and that it depends
only upon the quantities ${M}$ and $\Omega$ which are independent of any
choice of coordinates.

The second dimensionless parameter ${v_{12}^2}/{c^2} $ is known in terms of
$\gamma$ at the 2PN order to be
\begin{eqnarray}
  \frac{v_{12}^2}{c^2} =  \gamma - (3-\nu)\gamma^2
             + \Big(6 + \frac{41}{4}\nu +  \nu^2\Big) \gamma^3 + O(8)
\label{v12}
\end{eqnarray}
as given in BFP Eq.~(8.6) or Eq.~(190) in Ref.~\cite{BlanchetLR06}, with
$\nu = {m_1}{m_2}/{M}^2$.
 From \Deqns{xdef} and (\ref{v12})
\begin{equation}
     x^3 = \gamma^3 - (3-\nu)\gamma^4
             + \Big(6 + \frac{41}{4}\nu +  \nu^2\Big) \gamma^5 +O(12) .
\end{equation}
Alternatively, $\gamma$ may be expanded in terms of $x$ which results in
\begin{equation}
  \gamma = x + \Big(1-\frac{\nu}{3}\Big)x^2
               + \Big(1-\frac{65\nu}{12}\Big)x^3 +O(8)
\label{gammaEq}
\end{equation}
as in Eq.~(193) of ref.~\cite{BlanchetLR06}.

Now we focus on $u^t$ and its post-Newtonian expansion via the regularized
metric at ${m_1}$ given by BFP in their Eq.~(7.6).

Let $V^a = (c, \vec v_1)$, and let $u^a$ be parallel to $V^a$ but normalized
so that $u^au_a = -1$. Thus, $u^a = u^t V^a$ is the appropriately
parameterized four velocity of ${m_1}$.
 In this Appendix our major task is to find the post-Newtonian expression for
\begin{equation}
  u^t = 1/\sqrt{-V^a V^b g_{ab}}
      = 1/\sqrt{|g_{tt} + 2v_1^ig_{it} + v_1^i v_1^j g_{ij}|} .
\end{equation}

Our restriction to circular orbits allows for simplifications of the more
general regularized 2PN metric at the location of ${m_1}$ given in BFP
Eq.~(7.6).
 For example, $(n_{12}v_{12}) \equiv \vec r_{12}\cdot\vec v_{12}/r_{12} =
O(5)$, and such terms in Eq.~(7.6) do not lead to a contribution to $u^t$ at
O(6).
 Starting with BFP Eq.~(7.6), specializing to the circular orbit case, and
dropping terms whose contributions to $u^t$ are smaller than $O(6)$, we
obtain the metric components
\begin{widetext}
\begin{eqnarray}
 (g_{00})_1 &=&-1+2 \frac{G {m_2}}{c^2r_{12}}
       + \frac{G {m_2}}{c^4 r_{12}} \left(4 v_2^2 
       - 3\frac{G {m_1}}{r_{12}}- 2 \frac{G {m_2}}{r_{12}} \right)
   +
       \frac{G {m_2}}{c^6 r_{12}} \left( 
       4 v_2^4 \right)
      + \frac{G^2 {m_1} {m_2}}{c^6 r_{12}^2} \left(
      \frac{23}{4} v_1^2 - \frac{39}{2} (v_1v_2) \right)
\nonumber \\
      &+& \frac{47}{4} \frac{G^2 {m_1} {m_2}}{c^6 r_{12}^2} v_2^2+
\frac{G {m_2}}{c^6 r_{12}} \left\{ \frac{G {m_2}}{r_{12}}
  \left[ 
  - v_2^2 \right] - \frac{G^2 {m_1}^2}{r_{12}^2}
  + \frac{17}{2} \frac{G^2 {m_1} {m_2}}{r_{12}^2}
  + 2 \frac{G^2 {m_2}^2}{r_{12}^2}  \right\}
  +O(8) , \label{7.6a}
\end{eqnarray}
\begin{eqnarray}
(g_{0i})_1 &=&-4 \frac{G {m_2}}{c^3 r_{12}} v_2^i
 + \frac{G {m_2}}{c^5 r_{12}} \left\{
      4 \frac{G {m_1}}{r_{12}} v_1^i + v_2^i \left( 
    - 4 v_2^2- 2 \frac{G {m_1}}{r_{12}} + \frac{G {m_2}}{r_{12}} \right) \right\}
   +O(7) \ ,
\label{7.6b}
\end{eqnarray}
\begin{eqnarray}
(g_{ij})_1 &=& \delta^{ij} + 2 \frac{G {m_2}}{c^2 r_{12}} \delta^{ij}
        + \frac{G{m_2}}{c^4 r_{12}} \delta^{ij} \left( 
          \frac{G {m_1}}{r_{12}} + \frac{G {m_2}}{r_{12}} \right)
   + \frac{G {m_2}}{c^4 r_{12}} \left\{
    4 v_2^{i}v_2^{j}\right\}
  +O(6) \ . \label{7.6c}
\end{eqnarray}
\end{widetext} The order of terms, bracketing and other details in these
equations are as close as possible to the original form given in BFP to
facilitate a comparison of this form to the original, which an enthusiastic
reader might attempt.

In the center of mass frame of reference  
\begin{equation}
  v_1 \equiv |\vec v_1| = [{m_2}+3\gamma^2\nu({m_1}-{m_2})]\,v_{12}/{M} +O(6),
\label{v1eq}
\end{equation} 
and
\begin{equation}
  v_2 \equiv |\vec v_2| = [{m_1}-3\gamma^2\nu({m_1}-{m_2})]\,v_{12}/{M} +O(6),
\label{v2eq}
\end{equation} 
from Eq.~(187) of Ref.\ \cite{BlanchetLR06}.

With these substitutions, the use of \Deqns{7.6a}--(\ref{7.6c}) and $G$
being removed in favor of $\gamma$ via \Deqn{gammadef}, we obtain
\begin{widetext}
\begin{eqnarray}
  1/{u^t}^2 = - V^a V^b g_{ab} &=&
        1- \frac{{m_2}^2v_{12}^2}{{M}^2c^2} - 2 {\frac{{m_2}\gamma}{{M}}}
  -2{\frac{{m_2}^3{v_{12}}^2 \gamma}{{{M}}^3{c}^2}}\left(1 +
           4\mu_1 + 2\mu_1^2\right)
      + {\frac{{m_2}^2  {\gamma}^2}{{M}^2}}\left(2 + 3\mu_1\right)
   - 4\frac{{m_2}^3 \mu_1^2 v_{12}^4\gamma}{{M}^3c^4}
 \nonumber\\&& {}
   - \frac{{m_2}^4 v_{12}^2 \gamma^2}{4{M}^4 c^2}
        \left(4 + 27 \mu_1  + 114 \mu_1^2 + 47 \mu_1^3 \right)
   + \frac{{m_2}^3 {\gamma}^3}{2{{M}}^3}\left(- 4 - 17\mu_1 + 2\mu_1^2\right)
   + O(8),
\end{eqnarray}
with $\mu_1 \equiv {m_1}/M$.

Next $v_{12}^2/c^2$ is removed in favor of $\gamma$ using \Deqn{v12},
\begin{eqnarray}
  1/{u^t}^2 &=& 1 + \left(
  - 3
  + 4 \mu_1
  - \mu_1^2
  \right) \gamma
   + \left(
     3
   - 10 \mu_1
   + 11 \mu_1^2
   -  5 \mu_1^3
   +    \mu_1^4
   \right) \gamma^{2}
\nonumber\\ && {}
  + \left(
  - 3
  +  \frac{1}{2}  \mu_1
  +  \frac{1}{2}  \mu_1^2
  -  \frac{51}{4} \mu_1^3
  -  \frac{59}{4} \mu_1^4
  +  6            \mu_1^5
  -               \mu_1^6
  \right) \gamma^3 + O(8) .
\end{eqnarray}
And $\gamma$ is removed in favor of $x$ using \Deqn{gammaEq} which yields
\begin{eqnarray}
  u^t &=& 1
 + \left(
              \frac{3}{2}
          - 2 \mu_1
          + \frac{1}{2} \mu_1^2
   \right) x
 + \left(
            \frac{27}{8}
          - \frac{13}{2}  \mu_1
          + \frac{53}{12} \mu_1^2
          - \frac{4}{3}   \mu_1^3
          + \frac{1}{24}  \mu_1^4
   \right) x^2
\nonumber\\ && {}
 + \left(
           {\frac{135}{16}}
          - {\frac{175}{8}}  \mu_1
          + {\frac{409}{16}} \mu_1^2
          - {\frac{97}{6}}   \mu_1^3
          + {\frac{69}{16}}  \mu_1^4
          - \frac{1}{4}      \mu_1^5
          - \frac{1}{48}     \mu_1^6
   \right) x^3
 + O(8)
 \label{utx}
\end{eqnarray}
\end{widetext}
as the O(6) expression for $u^t$ of ${m_1}$ in a circular binary with
${m_2}$.

To change from this expansion, appropriate for a comparable mass binary, to a
similar one with an extreme mass ratio it is necessary to replace the
post-Newtonian dimensionless parameter
\begin{equation}
   x  = \left[\frac{\Omega G {m_2}}{c^3}\left(1+\frac{{m_1}}{{m_2}}\right)\right]^{2/3}
\end{equation}
with a dimensionless parameter
\begin{equation}
   x_{\epsilon} \equiv \left(\frac{\Omega G {m_2}}{c^3}\right)^{2/3}
\end{equation}
more convenient for simultaneous, independent expansions in $v/c$ and in
$\epsilon\equiv{m_1}/{m_2}$. [Note that $x_\epsilon = Gm_2/R_\Omega c^2$,
from \Deqn{Romdef}.]
 It follows that
\begin{equation}
  x = x_{\epsilon} (1+\epsilon)^{2/3} ,
\end{equation}
and expanding yields
\begin{equation}
  x= x_{\epsilon} \left[
        1
      + \frac{2}{3}  \epsilon
      - \frac{1}{9}  \epsilon^2
      + \frac{4}{81} \epsilon^3
      + O(\epsilon^4)
      \right] .
\label{xepsilon}
\end{equation}
The final steps are the substitutions of $\mu_1= \epsilon/(1+\epsilon)$ and
of \Deqn{xepsilon} into \Deqn{utx} and the subsequent collection of powers
of the small quantities $\epsilon$ and of $x_\epsilon$. Thus, for an extreme
mass ratio with a particle of mass ${m_1}$ in a circular orbit with the
orbital frequency $\Omega$ about a Schwarzschild black hole of mass ${m_2}$
\begin{eqnarray}
 u^t &=& 1
  + \left(
     \frac{3}{2}
    - \epsilon
    + \epsilon^2
    - \frac{28}{27} \epsilon^3
    \right) x_{\epsilon}
 \nonumber\\ &&
  + \left(
        \frac{27}{8}
    - 2 \epsilon
    + 3 \epsilon^2
    - \frac{67}{18} \epsilon^3
  \right) x_{\epsilon}^2
\nonumber\\ &&
 {}+ \left(
        \frac{135}{16}
    - 5 \epsilon
    +   \frac{97}{8} \epsilon^2
    -   \frac{97}{6} \epsilon^3
    \right) x_{\epsilon}^3
 \nonumber\\ &&
  +O(8)+ O(2)\times O(\epsilon^4) .
\end{eqnarray}
This result is consistent with the post-Newtonian expansion of the geodesic
value of $u^t$
\begin{eqnarray}
  {}_0u^t&=& (1-3m/r)^{-1/2} = (1-3x_\epsilon)^{-1/2}
 \nonumber\\ &=& 1 + \frac{3}{2}x_\epsilon
       + \frac{27}{8} x^2_\epsilon + \frac{135}{16} x^3_\epsilon +
       O(x_\epsilon)^4 .
\end{eqnarray}
And the first order self-force effect on $u^t$ is
\begin{eqnarray}
  {}_1u^t&=&  -\epsilon x_\epsilon -2 \epsilon x_\epsilon^2 - 5\epsilon x_\epsilon^3
    + O(\epsilon x_\epsilon^4),
\label{PNexpanA}
\end{eqnarray}
which is plotted in Figs.~(\ref{20-150uT}) and (\ref{5-20uT}). In the
future, a second-order perturbation analysis might be compared with
\begin{eqnarray}
  {}_2u^t&=&  \epsilon^2 x_\epsilon
        + 3 \epsilon^2 x_\epsilon^2 + \frac{97}{8}\epsilon^2 x_\epsilon^3
    + O(\epsilon^2 x_\epsilon^4) .
\end{eqnarray}


\bibliographystyle{prsty}


\end{document}